\documentclass[prb,twocolumn,aps,pacs,superscriptaddress,amsmath,amssymb]{revtex4-1}

\usepackage{graphicx}% Include figure files
\usepackage{dcolumn}% Align table columns on decimal point
\usepackage{bm}
\usepackage{ulem}
\usepackage{epsfig}
\usepackage{color}
\usepackage{textcomp}
\usepackage[lofdepth,lotdepth,caption=false]{subfig}
\usepackage[breaklinks=true,colorlinks,citecolor=blue,linkcolor=blue,urlcolor=blue]{hyperref}

\begin{document}
\title{Nanoscale Mach-Zehnder interferometer \\ with spin-resolved quantum Hall edge states}

\author{Biswajit Karmakar}
\affiliation{Saha Institute of Nuclear Physics, I/AF Bidhannagar, Kolkata 700 064, India}
\affiliation{NEST, Scuola Normale Superiore, and Istituto Nanoscienze-CNR, I-56126 Pisa, Italy}

\author{Davide Venturelli}
\affiliation{NASA Ames Research Center Quantum Artificial Intelligence Laboratory (QuAIL), Mail Stop 269-1, 94035 Moffett Field CA}
\affiliation{USRA Research Institute for Advanced Computer Science (RIACS), 615 National, 94043 Mountain View CA}

\author{Luca Chirolli}
\affiliation{IMDEA Nanoscience, Calle de Faraday 9, E-28049, Madrid, Spain}

\author{Vittorio Giovannetti}
\affiliation{NEST, Scuola Normale Superiore, and Istituto Nanoscienze-CNR, I-56126 Pisa, Italy}

\author{Rosario Fazio}
\affiliation{ICTP, Strada Costiera 11, I-34151 Trieste, Italy}
\affiliation{NEST, Scuola Normale Superiore, and Istituto Nanoscienze-CNR, I-56126 Pisa, Italy}

\author{Stefano Roddaro}
\affiliation{NEST, Istituto Nanoscienze-CNR and Scuola Normale Superiore, I-56126 Pisa, Italy}

\author{Loren N. Pfeiffer}
\affiliation{School of Engineering and Applied Science, Princeton University, New Jersey, USA}

\author{Ken W. West}
\affiliation{School of Engineering and Applied Science, Princeton University, New Jersey, USA}

\author{Fabio Taddei}
\affiliation{NEST, Istituto Nanoscienze-CNR and Scuola Normale Superiore, I-56126 Pisa, Italy}

\author{Vittorio Pellegrini}
\affiliation{Istituto Italiano di Tecnologia, Graphene Labs, Via Morego 30, I-16163 Genova, Italy}
\affiliation{NEST, Istituto Nanoscienze-CNR and Scuola Normale Superiore, I-56126 Pisa, Italy}

\begin{abstract}
We realize a nanoscale-area Mach-Zehnder interferometer with co-propagating quantum Hall spin-resolved edge states and demonstrate the persistence of gate-controlled quantum interference oscillations, as a function of an applied magnetic field, at relatively large temperatures.
Arrays of top-gate magnetic nanofingers are used to induce a resonant charge transfer between the pair of spin-resolved edge states.
To account for the pattern of oscillations measured as a function of magnetic field and gate voltage, we have developed a simple theoretical model which satisfactorily reproduces the data.
\end{abstract}


\maketitle

\section{Introduction}
Quantum technologies are emerging as a precious outcome of our improved ability to manipulate and control coherent devices.
In the solid state, electronic quantum interferometers realised making use of the topological protection of dissipationless quantum Hall (QH) conducting edge states are an example~\cite{Ji03,Neder06,Roulleau07,Litvin0708,McClure09,Deviatov11,Deviatov12}.
One of the motivations driving research activity on such interferometers is the opportunity to use them as quantum sensing devices (such as charge and "which-path" detectors~\cite{Sprinzak00,Neder07,Weisz12}), as on-demand single-electron sources\cite{Feve07}, and, possibly, as quantum heat engines\cite{Sothmann14,Sanchez15,Hofer15,Sanchez15-2}.
In this paper we report the observation of coherent oscillations of electric current, persisting at relatively high temperatures, in a well-controlled Mach-Zehnder (MZ) interferometer of small area.

\begin{figure}[h!]
\centering
\includegraphics[width=0.5\textwidth]{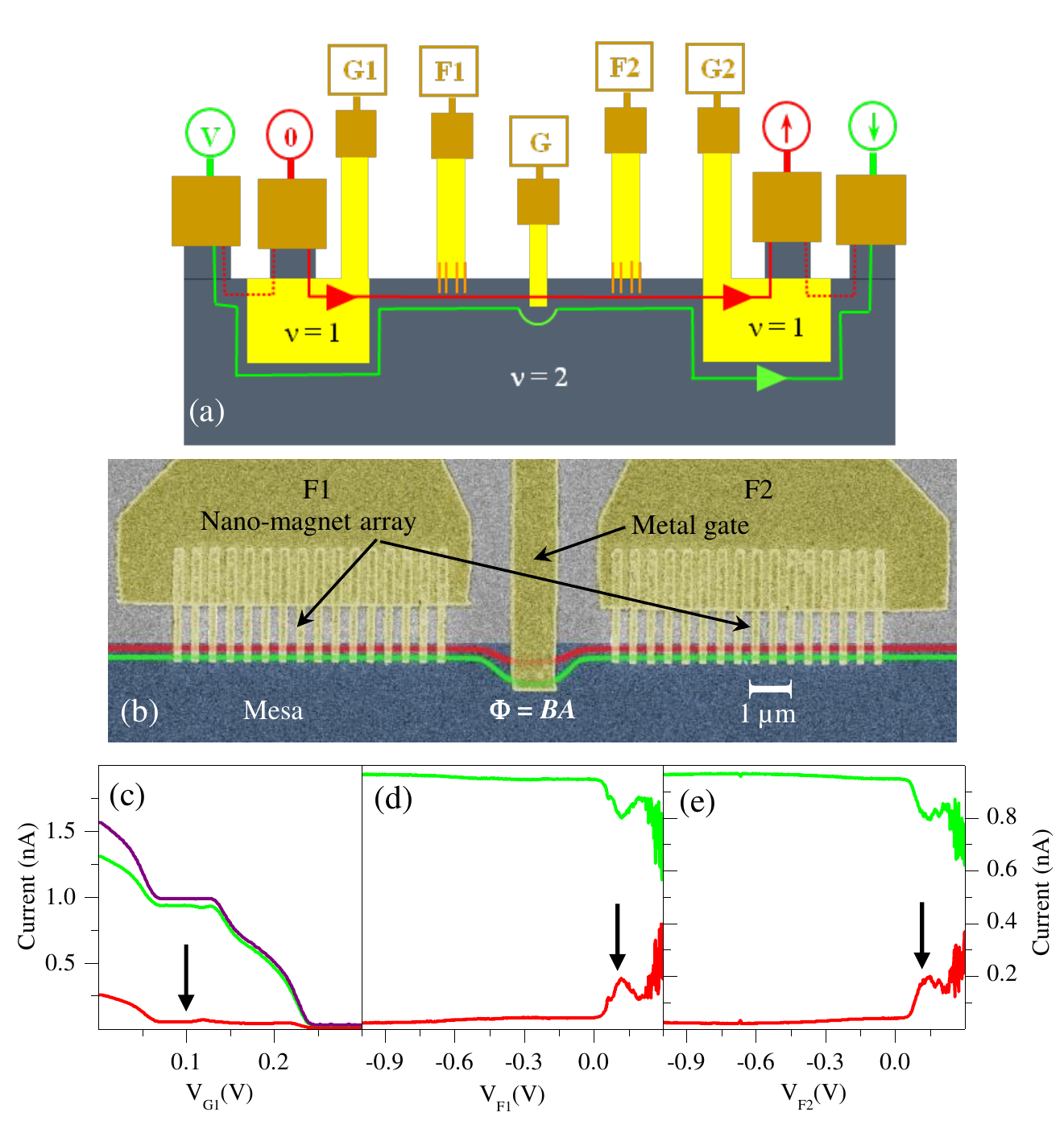}
\caption{(a) Schematic setup of quantum interference device with separately contacted co-propagating SRESs (red and green lines) at filling factor $\nu= 2$. G1, G2 and G denote gates and F1 and F2 denote magnetic nanofingers. (b) SEM shows the two arrays of magnetic nanofingers F1 and F2 and the metal gate G. Red and green lines schematically represent spin-up and spin-down SRES at the boundary of the mesa. (c) G1 gate bias dependence of output current shows the working point (black arrow) of 0.1 V, where input edge channels are separately contacted. In this measurement the output edge channels are separately contacted by applying $V_{\text{G2}} = 0.1$ V and the mixing of SRESs is deactivated by applying the following voltage biases on the nano fingers $V_{\text{F1}}=V_{\text{F2}}= -2$ V.
Red, green and maroon lines represent spin up, spin down channel current and total current respectively.
(d) F1 fringe bias dependence of magnetic coupling of SRESs shows the working point (black arrow) at $V_{\text{F}1}=0.09$ V, while $V_{\text{F}2}$ is kept at $-2$ V. (e) F2 fringe bias dependence of magnetic coupling of SRESs shows the working point (black arrow) at $V_{\text{F}2}=0.09$ V, while $V_{\text{F}1}$ is kept at $-2$ V.}
\label{fig1}
\end{figure}

In an optical MZ interferometer the two light beams produced by a beam splitter (BS), after having accumulated a controllable phase difference, recombine at a second BS where they interfere producing outgoing light beams whose amplitude oscillates as a function of the phase difference.
Refs.~\onlinecite{Ji03,Neder06,Roulleau07,Litvin0708} report on the realization of an electronic version of a MZ interferometer using counter propagating QH edge states arising in a clean two-dimensional electron gas (2DEG) under a high perpendicular magnetic field. In these devices the BSs are implemented through quantum point contacts (QPCs) of controllable transmission. The typical working temperature is of the order of a few tens of mK, while the area enclosed by the interfering paths is of the order of 50 $\mu$m$^2$.
On the other hand, co-propagating edge states are used in Refs.~\onlinecite{Deviatov11,Deviatov12} to implement a MZ interferometer. In these works the coupling between edge states (which implements a BS) of the same Landau level is induced by a large enough, i.e. far beyond equilibrium imbalance between edge states. Oscillations are observed up to relatively high temperatures, of the order of 0.5 K, with loop areas of the order of 0.1 $\mu$m$^2$.

In this paper we consider the device sketched in Fig.~\ref{fig1}a, which implements the MZ interferometer architecture proposed in Ref.~\onlinecite{Giovannetti08} and further analyzed in Ref.~\onlinecite{Karmakar13}, which is based on co-propagating edge states.
Here we work at filling factor (number of filled Landau levels) $\nu= 2$, i.e. with spin-resolved edge states (SRES).
In the scanning electron micrograph (SEM) of the structure, Fig.~\ref{fig1}b, red and green solid lines are drawn to represent outer (spin-up) and inner (spin-down) SRESs, respectively.
The MZ interferometer consists of two beam splitters (BSs), which couple the two co-propagating SRESs, separated by a region of area $A$ under the gate G where an Aharonov-Bohm (AB) flux $\Phi=BA$ is enclosed between the SRES ($B$ is the applied perpendicular magnetic field).
In our device each BS is realized with an array of top-gate magnetic nanofingers (F1 and F2 in Fig.~\ref{fig1}a and \ref{fig1}b), placed at the boundary of the 2DEG, which induces a resonant charge transfer between the two SRESs occurring when the array's periodicity compensates the momentum difference $\Delta k$ of the two SRES~\cite{Karmakar11}.
Such a local and controlled mixing between a pair of co-propagating SRES has been demonstrated in a previous experiment~\cite{Karmakar11}.
The MZ interferometer operation is realized by injecting electrons into the spin-down SRES from the left of Fig.~\ref{fig1}b, and measuring the current on the right in either SRES.
As a result of interference between the two SRES, the measured current is expected to oscillate as a function of $\Phi$. The latter can be varied by changing $B$ or by changing the voltage $V_G$ applied to the central gate G (see Fig.~\ref{fig1}b) that controls the area $A$.
Our measurements, performed in an equilibrium condition, show that in our device current oscillations occur up to relatively high temperatures, of the order of 0.5 K, with a small area between the interfering paths, of the order of 0.05 $\mu$m$^2$, which results in an enhanced sensitivity.
Our device is characterized, remarkably, by its very good controllability through the various gate voltages which, in particular, control the coupling between the two co-propagating SRES, i.e. the transmission of the BSs.

The paper is organized as follows: the experimental setup and the calibration of the device are described in Sec.~\ref{exp}, while Sec.~\ref{results} is devoted to the discussion of the result of the measurements, i.e. the output current of the interferometer. In Sec.~\ref{theo} a simple theoretical model which accounts for the data is described, while the conclusions are drawn in Sec.~\ref{conc}. Moreover, a discussion regarding a ``cross-over'' effect is included in the appendix.

\section{Experimental setup and calibration}
\label{exp}

The device is fabricated on a modulation-doped AlGaAs/GaAs heterostructure grown by molecular beam epitaxy. The 2DEG resides at the AlGaAs/GaAs heterointerface located 80 nm below the top surface. An undoped AlGaAs spacer layer of thickness 50 nm separates the 2DEG from the Si $\delta$-doping layer, which supply free electrons at the heterointerface. The 2DEG has nominal electron density of $3\times 10^{11}$/cm$^2$ and low-temperature mobility nearly $8\times 10^6$ cm$^2$/Vs. The interferometer device is fabricated by standard photo-lithography and electron beam lithography (EBL). Cobalt nanomagnet array in the device is defined at the mesa boundary using EBL and thermal evaporation of 10 nm Ti followed by 120 nm Co and then 10 nm of Au.
For electrical transport measurements the device is mounted on a He3 wet cryostat equipped with 12 Tesla superconducting magnet. In dark condition the sample does not conduct at low temperatures. It starts conducting after light illumination with infrared GaAs LED at temperature 4 K and then it is cooled down to 250 mK. Persistence photo conductivity holds the injected carriers in the 2DEG for long.

The sketch in Fig.~\ref{fig1}a shows how the two SRESs are separately contacted.
By tuning the voltage applied to the top gates G1 and G2, the filling factor beneath them can be lowered to $\nu=1$ so that only the outer SRES (red) passes beneath the gates and connects to a Ohmic contact.
Inner SRESs (green) are deflected around the gates and connected to the farthest Ohmic contacts.
On the left-hand-side the (injection) contact for the inner edge is kept at voltage $V$, while the contact for the outer edge is grounded.
Inner and outer SRESs currents are then measured on the right-hand-side (detection) contacts, denoted by $\uparrow$ and $\downarrow$.
In the SEM (Fig.~\ref{fig1}b) the dark blue region is the mesa defined by EBL and wet chemical etching. The two sets of magnetic nanofingers arrays are fabricated at the mesa boundary with an overlap of nearly 400 nm. An array periodicity of 400 nm is chosen from our previous experiments on a similar 2DEG~\cite{Karmakar11} to maximize SRESs mixing.
The central gate G has an overlap area of nearly $1\times1~\mu$m$^2$ on the mesa.
In order to control the SRESs mixing and activate/deactivate the BSs, both arrays are electrically contacted to Au pads where a voltage can be applied ($V_{\text{F}1}$ and $V_{\text{F}2}$).
Indeed, one expects that by applying a negative voltage $V_{\text{F}i}$ the portion of the 2DEG beneath the nanofingers is depleted and the SRESs move inward in the mesa, that is away from the region where the magnetic fringe field of the nanofingers is significant (see below).

Preliminary measurements and calibrations are performed as follows.
Initially the two terminal magneto resistance is measured to determine the magnetic field ($B = 5.56$ T) required to set the filling factor of the 2DEG to $\nu=2$.
Current measurements are carried out by standard lock-in techniques using current to voltage preamplifiers. An ac voltage of $V = 25.8~\mu$V at 17 Hz is applied on the inner edge channel (green line) contact denoted by V in Fig.~\ref{fig1}a and currents are measured at the terminals $\uparrow$ and $\downarrow$. 
The working point values of the voltages applied to gates G1 and G2 at which the SRESs are separately contacted are determined by measuring the current $I_{\uparrow}$, $I_{\downarrow}$ and their sum as functions of $V_{\text{G}1}$ and $V_{\text{G}2}$ as in Ref.~\onlinecite{Karmakar11}.
This measurement is performed by keeping the magnetic arrays deactivated with a large negative voltage ($V_{\text{F}1}=V_{\text{F}2}=-2$ V).
As an example, Fig.~\ref{fig1}c shows that the total transmitted current becomes zero at $V_{\text{G1}} > 0.25$ V, since filling factor $\nu=2$ is attained beneath the gate G1 and current is shunted through the neighbouring grounded Ohmic contact.
When the voltage $V_{\text{G1}} = 0.1$ V is reached, only the spin-up SRES is shunted (red dotted line in Fig.~\ref{fig1}a), while the spin-down SRES (green line) is transmitted going around the gate. Note that the total transmitted current is nearly equal to 1 nA (resulting from 25.8 $\mu$V voltage excitation divided by the resistance quantum, 25.8 k$\Omega$), i.e. the green channel carries nearly all the current.
This suggests that the spin relaxation between the co-propagating SRESs, in absence of the coupling of the nanofingers arrays, is very small on a distance of the order of 100 $\mu$m (namely, the separation between injection and detection contacts on the mesa).
As the voltage on the gate G1 is further reduced to zero, the transmitted currents increase. This enhancement of the transmitted currents is the signature of the reduction of the filling factor $\nu< 1$ beneath the gate G1.
When the G1 gate voltage is set to $-0.5$ V (not shown in Fig.~\ref{fig1}c), filling factor $\nu = 0$ is attained so that both SRESs are transmitted and the total current reaches 2 nA.

The mixing induced by the arrays of nanofingers is then tested by changing the voltage applied to one gate (say F1) while the other gate (F2) is kept at a large negative voltage to deactivate the SRESs mixing.
As shown in Fig.~\ref{fig1}d (Fig.~\ref{fig1}e), when $V_{\text{F}1}=0.09$ V ($V_{\text{F}2}=0.09$ V) and $V_{\text{F}2}=-2$ V ($V_{\text{F}1}=-2$ V), $I_{\uparrow}$ reaches a peak value of nearly 0.2 nA (the total current remaining constant at 1 nA).
Analogously to Ref.~\onlinecite{Karmakar11}, such SRESs mixing is consistent with a resonant charge transfer induced by the magnetic nanofingers arrays.

\section{Results}
\label{results}
\begin{figure}[t!]
\centering
\includegraphics[width=0.5\textwidth]{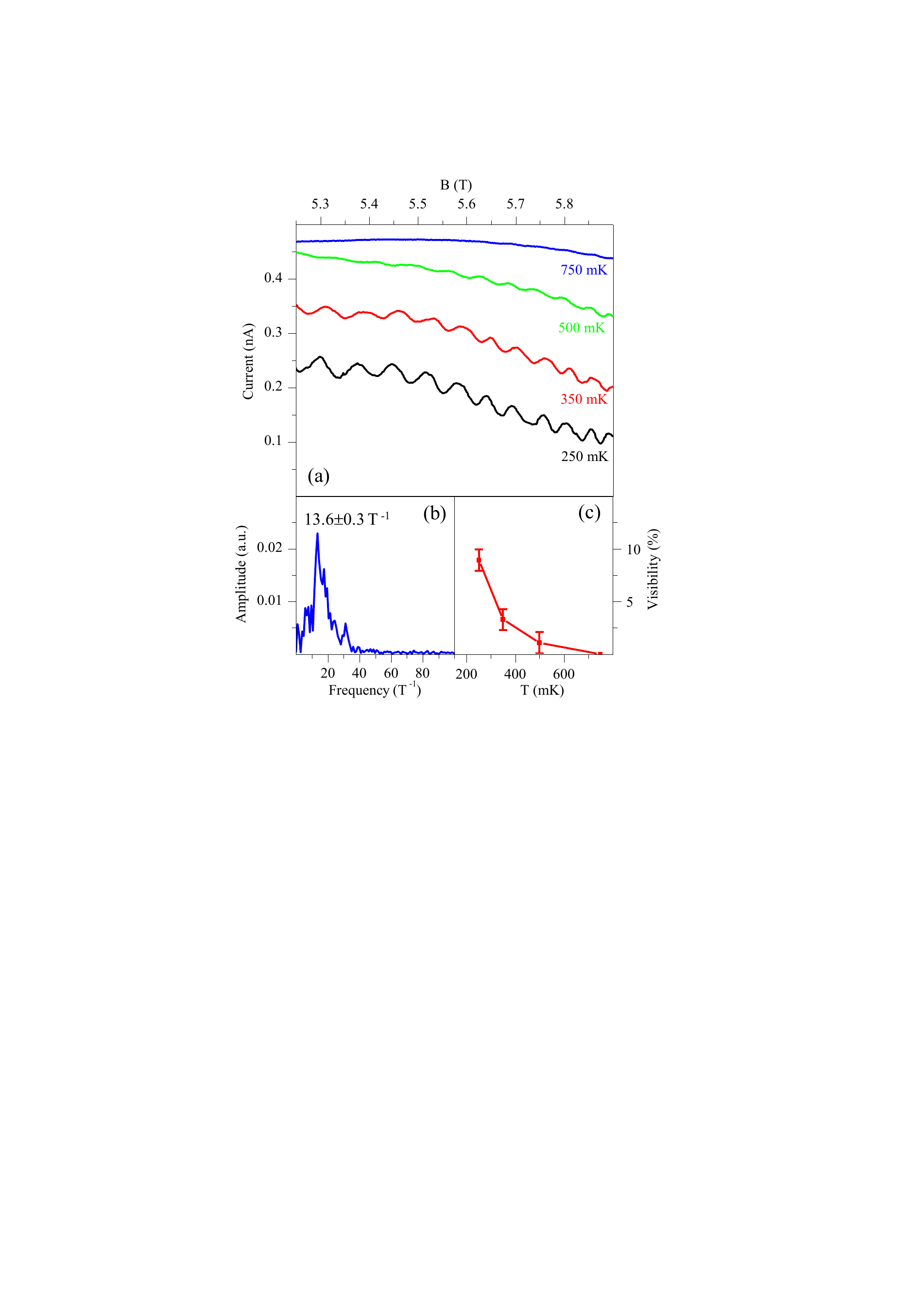}
\caption{(a) Transmitted spin-up current $I_{\uparrow}$ as a function of the magnetic field $B$ within the $\nu=2$ plateau for different temperatures. Gate voltages are set as follows: $V_G=0.35$ V, $V_{F1}=V_{F2}=0.09$ V. Interference oscillations weakens with increasing temperature $T$. (b) FFT of the oscillations of the current measured at 250 mK: the peak corresponds to the frequency $f=13.6\pm 0.6$ T$^{-1}$. (c) The visibility of the oscillations, plotted as a function of temperature, decreases from 10\% at 250 mK to 0 at 750 mK.}
\label{fig2}
\end{figure}

The MZ interferometer is operated by activating both BSs, i.e. applying to the nanofingers F1 and F2 a voltage of 0.09 V, and fixing the central gate voltage at $V_G = 0.35$ V.
In Fig.~\ref{fig2} quantum interference oscillations are shown when sweeping the magnetic field $B$ within the extent of the $\nu=2$ plateau from 5.25 T to 5.9 T.
The transmitted current $I_{\uparrow}$ is plotted for different temperatures, from 250 mK to 750 mK.
The curves present oscillations and an overall current decrease with $B$, which is due to the diminishing of the SRESs mixing at the nanofingers with increasing $B$.
Strikingly, the oscillations, though getting weaker with increasing temperature, still persist up to 500 mK.

To analyze the temperature dependence of the oscillations we define the visibility $\mathcal{V}$ as the relative difference between maximum and minimum values of current after renormalizing with the low frequency component, to get rid of the inessential overall decrease of the signal with increasing B~\cite{Ji03}.
The visibility, plotted in Fig.~\ref{fig2}c as a function of temperature $T$, starts from around 10\% at 250 mK and decreases thereafter, indicating a decrease of coherence length $l_{\phi}$~\cite{Levkivskyi08}.
More precisely, we fit the curve in Fig.~\ref{fig2}c using the expression~\cite{Hashisaka10}
\begin{equation}
\mathcal{V}\propto \left(
1+\frac{T}{T_0} \right)
e^{-\frac{T}{T_0}} ,
\end{equation}
where $T_0$ is a fitting parameter taking the value $T_0\simeq 94$ mK, consistently with the results reported in Refs.~\onlinecite{Yamauchi09,Hashisaka10}.
The reason why interference persists at relatively high temperatures (note that the oscillations in the experiment of Ref.~\onlinecite{Ji03} disappear completely above 100 mK) has to be found in the fact that the loop area $A$ is small in our interferometer.
$A$ can be estimated from the frequency of oscillations $f$ which can be determined by Fourier transforming the current vs. field $B$ characteristic relative to 250 mK, see Fig.~\ref{fig2}b.
The main peak is found at $f = 13.6\pm 0.6$ T$^{-1}$, which corresponds to a AB loop area of $A = \phi_0 f = 0.056\pm 0.002$ $\mu$m$^2$, where $\phi_0=h/e=4.14 \times 10^{-15}$ Tm$^2$ is the flux quantum (in Ref.~\onlinecite{Ji03} the loop area was of the order of 50 $\mu$m$^2$).
The value of the area obtained is consistent with the geometry of the interferometer.
Indeed, since the gate G width is of the order of 1 $\mu$m, the average distance between the SRESs in the region between the BSs comes out to be of the order of a few tens of nanometers, consistently with a separation of a few magnetic lengths.
We have checked that the frequency is temperature-independent within the error of the measurement.
Finally note that the average value of the current increases by increasing temperature, reflecting the overall enhancement of spin relaxation leading to the mixing of the co-propagating SRESs.
\begin{figure}[th!]
\centering
\includegraphics[width=0.5\textwidth]{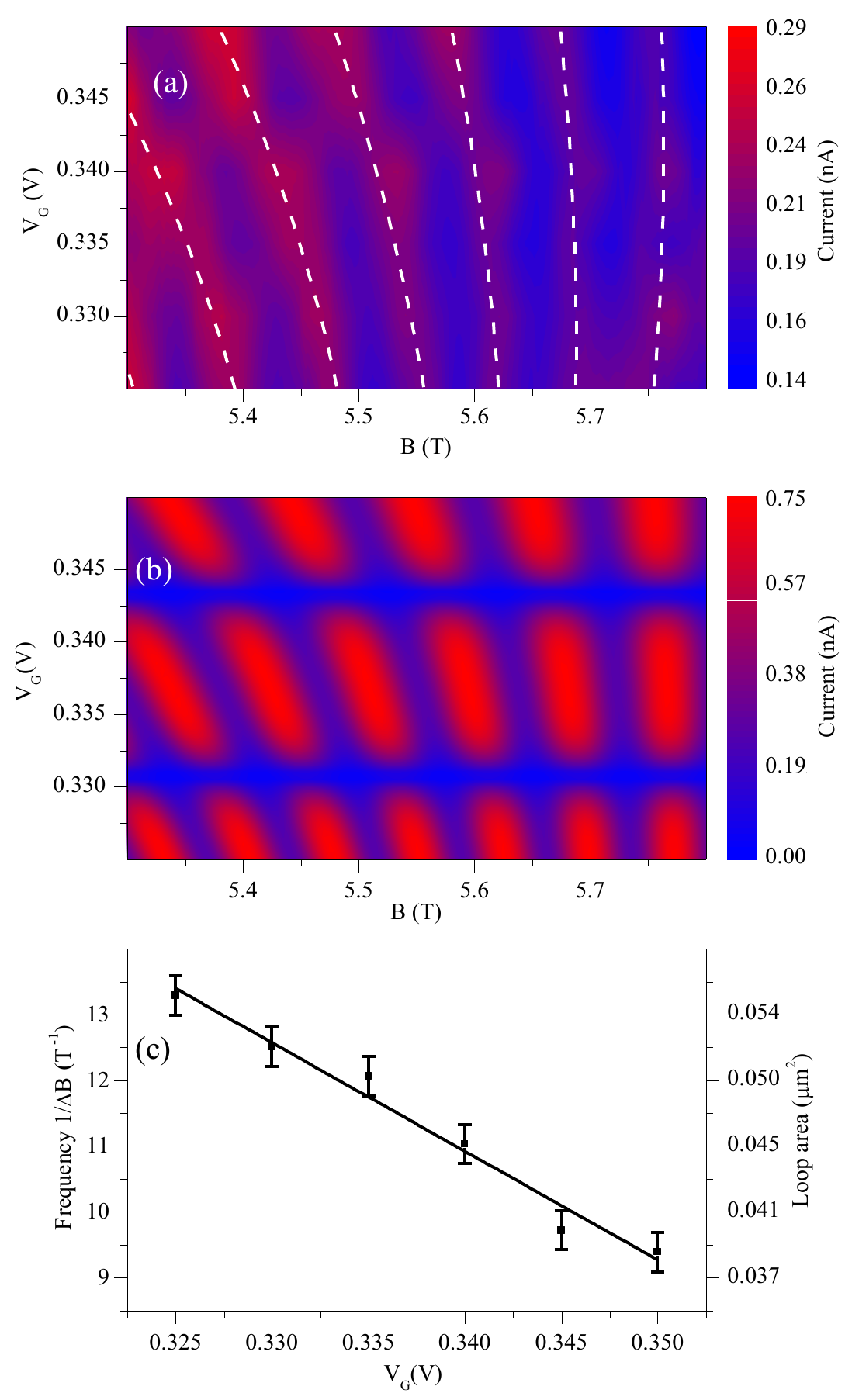}
\caption{(a) Color plot of the measured spin-up current $I_{\uparrow}$ showing the modulation through $V_G$ of the interference oscillations in $B$. (b) Color plot of the transmission probability for spin-up electrons calculated through a model which takes into account the ``cross-talk" effect of the voltage $V_G$ on the transmission properties of the arrays of nanofingers, and reproduces the discontinuous features. (c) Frequency of AB oscillations (left axis) and AB loop area (right axis) as a function of gate voltage $V_G$.}
\label{fig3}
\end{figure}

The AB loop area of the interferometer device can be modified by changing the voltage $V_G$ applied to the central gate.
The color plot in Fig.~\ref{fig3}a shows the transmitted current $I_{\uparrow}$ as a function of $B$ and $V_G$ at 250 mK, still setting F1 and F2 gate voltages to 0.09 V.
The pattern of oscillations exhibits discontinuous features and an overall decrease of the frequency of oscillations for increasing gate voltage, consistently with a decrease of the loop area when the voltage becomes more positive (the edge channels tend to move toward the boundary of the mesa).
The result of the Fourier transform of the oscillations is plotted in Fig.~\ref{fig3}c, which shows the frequency $f$ and the loop area $A$ as a function of $V_G$.
The discontinuous features in the pattern of oscillations in Fig.~\ref{fig3}a has been successfully reproduced with a model, described in the next section, that takes into account the influence of $V_G$ on the transmission properties of the BSs, the latter arising from the ``cross-talk" effect between the gates G and F1 and F2  (see Appendix~\ref{cross}).
The results of the calculations are shown in Fig.~\ref{fig3}b.

\section{Theoretical model}
\label{theo}
\begin{figure}[t]
\includegraphics[width=0.4\textwidth]{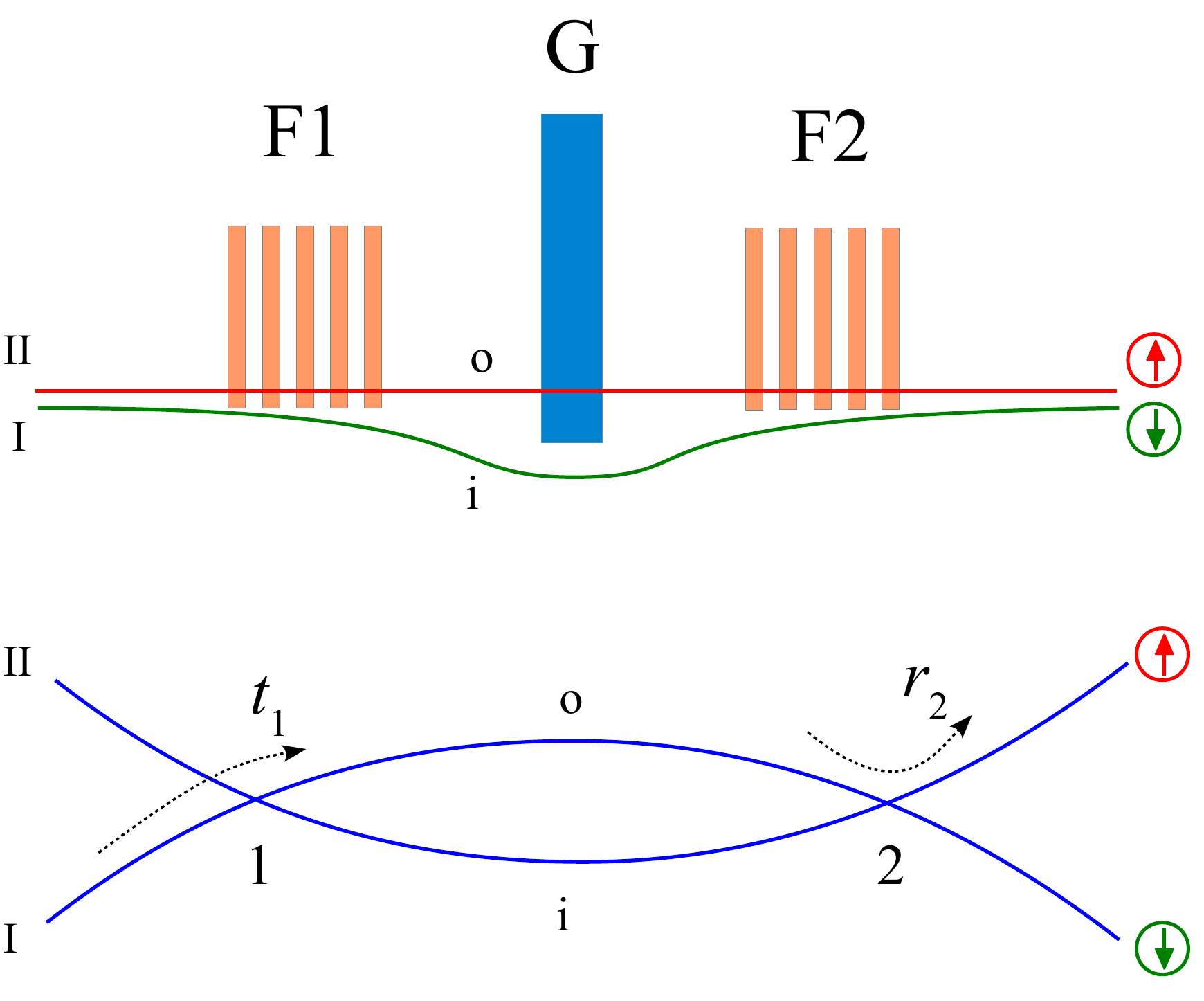}
\caption{Sketch of the MZ interferometer. Electrode I is biased at voltage $V$, while electrode II is grounded.}
\label{setup}
\end{figure}
The MZ interferometer consists of two BSs, 1 and 2, which couple the two edge state channels, inner (spin-down) in green and outer (spin-up) in red (see Fig.~\ref{setup}).
Electrode I is biased at voltage $V$, while all other electrodes are grounded.
Our aim is to study how the current $I_{\uparrow}$ measured in the electrode denoted by $\uparrow$ depends on magnetic field and gate voltage $V_\text{G}$.
The current $I_{\uparrow}$ is determined by the transmission probability $T_{\uparrow}=t_{\uparrow}^\star t_{\uparrow}$, through the Landauer-B\"uttiker scattering formula in the linear response regime
\begin{equation}
I_{\uparrow}=\frac{e^2}{h} T_{\uparrow} V .
\end{equation}
Note that, since the experimental data do not show dependence on the bias voltage, we have neglected the energy-dependence of the scattering amplitudes.
The transmission amplitude is given by
\begin{equation}
\label{t}
t_{\uparrow} = r_{2} e^{i\left( k_{\text{o}}L_{\text{o}}+\frac{e}{\hbar}\int_{\text{o}} \vec{A}\cdot d\vec{l} \right)} t_1 +
t_{2} e^{i\left( k_{\text{i}}L_{\text{i}}+\frac{e}{\hbar}\int_{\text{i}} \vec{A}\cdot d\vec{l} \right)} r_1 .
\end{equation}
In Eq.~(\ref{t}) $t_1$ and $t_{2}$ are the transmission amplitudes of the BS 1 and 2, while $r_1$ and $r_{2}$ are the corresponding reflection amplitudes.
Note that $t_1$ and $r_1$ depend on $V_{\text{F1}}$ and that $t_{2}$ and $r_{2}$ depend on $V_{\text{F2}}$.
It is however reasonable that all those scattering amplitudes also depend on $V_\text{G}$ (as a consequence of the``cross-talk" effect, see Appendix~\ref{cross}).
$k_{\text{o}}$ and $k_{\text{i}}$ are the wave vectors of the two edge channels, which are given by~\cite{Buttiker2005} $k_{\text{o}(\text{i})}=E/(\hbar v_{D,\text{o}(\text{i})})+k_{F,\text{o(i)}}$ [the energy $E$ is measured from the Fermi energy, $k_{F,\text{o(i)}}$ is the wave vector at the Fermi energy for the outer (inner edge state), and $v_{D,\text{o}(\text{i})}$ is the drift velocity].
The transmission probability can be written as
\begin{equation}
T_{\uparrow} =|r_{2}|^2 |t_{1}|^2 + |t_{2}|^2 |r_{1}|^2 + 2\Re \left[  r_{2}^\star t_{2}  t_{1}^\star r_{1} e^{i\Delta\Phi} \right] ,
\label{T41}
\end{equation}
where
\begin{equation}
\Delta\Phi=\frac{2\pi e}{h}\Phi + \Phi_{dy} ,
\end{equation}
\begin{equation}
\Phi_{dy}=(k_{\text{i}} L_{\text{i}}-k_{\text{o}} L_{\text{o}}) ,
\end{equation}
is the dynamical phase and
\begin{equation}
\Phi = \int_{\text{i}} \vec{A}\cdot d\vec{l} - \int_{\text{o}} \vec{A}\cdot d\vec{l} = BA
\end{equation}
is the magnetic flux (with magnetic field $B$) through the area $A$ enclosed by the two edge channels between the BSs.
Note that $A$ and $\Phi_{dy}$ (through $L_{\text{i}}$ and $L_{\text{o}}$) depend on $V_\text{G}$.
By setting
\begin{equation}
r_{2}^\star t_{2}  t_{1}^\star r_{1} = \tau e^{i\sigma},
\label{tau}
\end{equation}
with $\tau$ and $\sigma$ real numbers, Eq.~(\ref{T41}) can be rewritten as
\begin{equation}
T_{\uparrow} =|r_{2}|^2 |t_{1}|^2 + |t_{2}|^2 |r_{1}|^2 + 
2 \tau \cos \left( \Delta\Phi +\sigma \right) .
\label{T41-2}
\end{equation}

We shall furthermore assume that the parameters $A$, $L_{\text{o}}$, $L_{\text{i}}$, $k_{\text{o}}$, $k_{\text{i}}$ do not depend significantly on the magnetic field.
From now on let us fix $V_{\text{F1}}$ and $V_{\text{F2}}$ and consider only the dependence on $B$ and $V_\text{G}$.
According to Refs.~\onlinecite{Karmakar11,Chirolli2012}, the transmission amplitude of each set of nanofingers is a damped oscillating function (more precisely, a cardinal sine function) of the fingers' gate voltage $V_{\text{F}}$.
Indeed, a periodic modulation of the transmitted current through a single array of nanofingers is present in Figs.~\ref{fig1}d and \ref{fig1}e (though not clearly visible on this scale) around the working points marked by black arrows.
Because of the "cross-talk" effect, we now assume that both $t_1$ and $t_2$ depend on $V_\text{G}$ and take, for simplicity,
\begin{equation}
T_1=|t_1|^2=a_1 |\sin (\beta_1 V_\text{G} + \delta_1) |
\end{equation}
and
\begin{equation}
T_{2}=|t_2|^2=a_{2} |\sin (\beta_{2} V_\text{G} + \delta_{2}) |.
\end{equation}

If we now assume that both the area and the dynamical phase $\Phi_{dy}$ depend linearly on $V_\text{G}$, Eq.~(\ref{T41-2}) can be rewritten as
\begin{eqnarray}
\label{eqosc}
T_{\uparrow}&=& T_1 R_1+T_{2} R_{2} + \\
&&+\sqrt{T_1 R_1} \sqrt{T_{2} R_{2}} \cos [\frac{e}{\hbar}B (A_0-\alpha V_\text{G})+\gamma V_\text{G}] , \nonumber
\end{eqnarray}
where $\alpha>0$ so that the area $A=A_0-\alpha V_\text{G}$ decreases with (positive) $V_\text{G}$, and $R_i=1-T_i$, with $i=1,2$.
The parameters $A_0$, $\alpha$ and $\gamma$ can be inferred from the experimental data by matching the values of the angular frequencies of oscillations in $B$ and $V_\text{G}$ (Fig.~\ref{fig3}a).
We find
\begin{eqnarray}
\alpha=  \frac{\hbar}{e} 1032 \text{~m}^2\text{V}^{-1} \\
A_0=  \frac{\hbar}{e} 421 \text{~m}^{2} \\
\gamma=6000 \text{~V}^{-1},
\end{eqnarray}
where $\frac{\hbar}{e}=6.586\times10^{-16}$ Tm$^2$.
Note that the loop area $A$, at $V_\text{G}=0.34$ V, is of the order
\begin{equation}
A =A_0-\alpha V_\text{G} = 0.046 ~\mu\text{m}^2 .
\end{equation}
Fig.~\ref{fig3}b is obtained from Eq.~(\ref{eqosc}) using the above parameters and $a_1=a_2=0.5$, $\beta_1=\beta_2=250$ V$^{-1}$ and $\delta_1=\delta_2=-1$.
This shows that the inclusion of a ``cross-talk" effect, meaning that $V_G$ acts on the transmission properties of both BSs in the same way as $V_{F1}$ and $V_{F2}$, and including the dependence of $V_G$ on the loop area and the two paths forming the loop, is important to account for the experimental observations.
Indeed, the pattern of oscillations in the two figures (\ref{fig3}a and \ref{fig3}b) matches satisfactorily, even without introducing in the model the overall decrease of the current with increasing $B$.

\section{Conclusions}
\label{conc}
In conclusion, we have fabricated a MZ interferometer with scalable architecture where interference oscillations persist at relatively high temperatures (of the order of 0.5 K) making use of co-propagating, spin-resolved QH edge states.
Our device is characterized by a small interfering-paths area and by its very good controllability through the numerous gates which determine the transmission of the beam splitters and the separation between edge states in the various regions of the 2DEG.
To account for the peculiar features of the observed pattern of oscillations as a function of the gate voltages and magnetic field we have developed a theoretical model whose results agree satisfactorily with the data.

We note that the use of spin-resolved edge states is advantageous for several reasons.
On the one hand, the spatial separation between the two edge states is smaller with respect to spin-unresolved orbital edge states, and can be as small as a few nanometers~\cite{Chirolli2012}.
This implies a small-area interference loop of co-propagating edge states, ensuring a weaker impact of electromagnetic fluctuations of random origin, and an interference loop of small length which implies that electron transport remains coherent over a larger range of temperatures.

On the other hand, our QH spin-resolved interferometry is expected to be particularly suited for the integration with spintronics devices and for quantum information applications (see for example the scalable architecture of Ref.~\onlinecite{Yamamoto12}). For instance, a quantum bit (qubit) of information can be encoded in the spin degree of freedom, while operation on the qubit and coupling between qubits can be realized through top gates which affect the path of edge states. The readout is performed by measuring the current flowing through the contacts.

\section*{Acknowledgements}
This work has been supported by MIUR through FIRBIDEAS project No. RBID08B3FM and through PRIN project ``Collective quantum phenomena: from strongly correlated systems to quantum simulators".

\appendix
\section{``Cross-talk" effect}
\label{cross}
\begin{figure}[t!]
\centering
\includegraphics[width=0.4\textwidth]{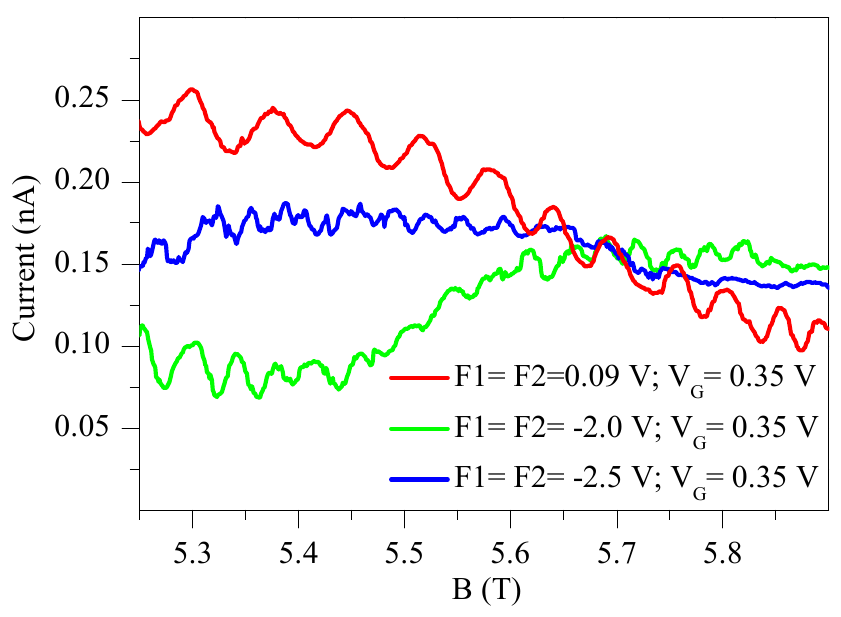}
\caption{``Cross-talk" effect: the gates F1 and F2 not only affects locally the strength of the coupling between SRESs, but also change the area $A$ of the interference loop. Spin-up current $I_{\uparrow}$ as a function of the magnetic field $B$ within the $\nu=2$ plateau for fixed $V_\text{G}=0.35$ V and different values of gate voltages $V_{\text{F1}}$ and $V_{\text{F2}}$. Data are taken at 250 mK.}
\label{fig4}
\end{figure}

The ``cross-talk" effect between the gates G and F1, and G and F2 arises because the distance between gate G and the two nanofingers' arrays is small (about of 1 $\mu$m, see the SEM micrograph in Fig.~\ref{fig1}b.
As a consequence, the voltage applied to the gates F1 and F2 not only affects locally the strength of the coupling between SRESs, but also changes the area $A$ of the interference loop.
This effect is demonstrated in Fig.~\ref{fig4} where we plot the spin-up current $I_{\uparrow}$ as a function of the magnetic field $B$ for different values of $V_{\text{F1}}$ and $V_{\text{F2}}$, while keeping fixed $V_\text{G} = 0.35$ V.
The red curve (taken from Fig.~\ref{fig2}a) is a reference showing oscillations when the beam splitters (BSs) are active ($V_{\text{F1}} =V_{\text{F2}} = 0.09$ V and $V_G = 0.35$ V).
By setting $V_{\text{F1}} =V_{\text{F2}}=-2$ V (green curve) we still clearly observe oscillations (although less regular, of reduced visibility and faster, as for a slightly larger loop area).
This behavior clashes strikingly with the plot in Figs.~\ref{fig1}d (and also with the plot in Figs.~\ref{fig1}e), whereby the spin-up current is totally suppressed already at $V_{\text{F1}} =-0.9$ V and $V_{\text{F2}}=-2$ V.
The difference is made by the fact that while the data in Fig.~\ref{fig4} were taken at $V_\text{G} = 0.35$ V, the data in Figs.~\ref{fig1}d and ~\ref{fig1}e were taken at $V_\text{G} = -1$ V.
Indeed, the attractive positive voltage applied to G ($V_\text{G}=0.35$ V) keeps the SRESs close to the nanofingers' arrays, at least in some portion of the latter in proximity to G, thus keeping finite the coupling between the SRESs (the beam splitter is active).
This indicates that the SRESs continue to be, at least, partially exposed to the nanofingers fringe field.
By further decreasing the nanofingers' gate voltage at $V_{\text{F1}} =V_{\text{F2}} = -2.5$ V (blue curve in Fig.~\ref{fig4}), we still get a finite current with oscillations, although of poorer quality.

Another consequence of the ``cross-talk" effect, discussed in Sec.~\ref{theo}, is that $V_\text{G}$ not only controls the position of the two SRESs (i.e. their lengths and the loop area $A$), but also influences the coupling between SRES.
We stress that the ``cross-talk" effect is unwanted and can be removed by increasing the separation between the gate G and the nanofingers.

\end{document}